\documentclass{ws-procs975x65}

\begin{document}

\title{THE IMPORTANCE OF BEING INTEGRABLE:\\ OUT OF THE PAPER, INTO THE LAB}

\author{MURRAY T. BATCHELOR}

\address{Centre for Modern Physics, Chongqing University,\\
Chongqing 400044, China\\ 
Australian National University, Canberra, ACT 0200, Australia\\
E-mail: Murray.Batchelor@anu.edu.au}

\begin{abstract}
The scattering matrix ($S$-matrix), relating the initial and final states of a physical system 
undergoing a scattering process, is a fundamental object in quantum mechanics and quantum field theory. 
The study of factorized $S$-matrices, in which many-body scattering factorizes into a product of two-body 
terms to yield an integrable model, has long been considered the domain of mathematical physics. 
Many beautiful results have been obtained over several decades for integrable models of this kind, 
with far reaching implications in both mathematics and  theoretical physics. The viewpoint that these 
were only toy models changed dramatically with brilliant experimental advances in realizing 
low-dimensional quantum many-body systems in the lab. These recent experiments involve both 
the traditional setting of condensed matter physics and the trapping and cooling of atoms in optical 
lattices to engineer and study quasi-one-dimensional systems. 
In some cases the quantum physics of one-dimensional systems is arguably more interesting than their 
three-dimensional counterparts, because the effect of interactions is more pronounced when atoms 
are confined to one dimension. This article provides a brief overview of these ongoing developments, 
which highlight the fundamental importance of integrability.
\end{abstract}


\bodymatter

\section{Introduction}\label{intro}
The goal of theoretical physics is to develop theories for the physical description of reality.
This provides ample enough motivation to study model systems which are constructed to 
capture the essential physics of a given problem. 
The strong predictive power of such basic models is one of the triumphs of theoretical physics.
On the other hand, mathematical models of this kind often become interesting in their own right, leading  
into the realms of mathematical physics. 
If the mathematical structures are sufficiently rich then progress can be inspired in mathematics itself.

The particular models we have in mind here are the so-called integrable models of 
statistical mechanics and quantum field theory. 
Their origin dates back to soon after the development of quantum mechanics, 
when the eigenspectrum of the one-dimensional spin-$\frac12$ Heisenberg chain 
was obtained in exact closed form by Hans Bethe.\cite{Bethe}
The underlying Bethe Ansatz for the wave function is the hallmark of the integrable models to be discussed here.
Indeed, these models can be referred to as being Bethe Ansatz integrable.
Some key examples from a golden period in the 1960's when the inner workings of the 
one-dimensional models were uncovered are:

\begin{itemize}

\item Bose gas, Lieb \& Liniger\cite{Lieb}, McGuire\cite{McGuire1}, Berezin {\em et al.}\cite{BPF} (1963,1964)

\item Fermi gas, $M=1$, McGuire\cite{McGuire2} (1965) 
      
\item Fermi gas, $M=2$, Flicker \& Lieb \cite{Flicker} (1967) 

\item Fermi gas, $M$ arbitrary, Gaudin\cite{Gaudin,Gaudin_thesis}, Yang\cite{Yang1,Yang2} (1967,1968) 

\item Fermi gas, higher spin, Sutherland\cite{Sutherland_fermions} (1968) 

\item Hubbard model, Lieb \& Wu\cite{Wu} (1968) 

\end{itemize}
Here $M$ is the number of spins flipped from the ferromagnetic state.

Another strand of developments during this golden period -- later seen to be not unrelated -- was sparked by Lieb's exact solution 
of the ice-type models, which culminated in Baxter's invention of the commuting transfer matrix and functional equation 
method to solve the eight-vertex model.\cite{Baxter}

\begin{figure}[h]
\begin{center}
\includegraphics [width=0.98\linewidth]{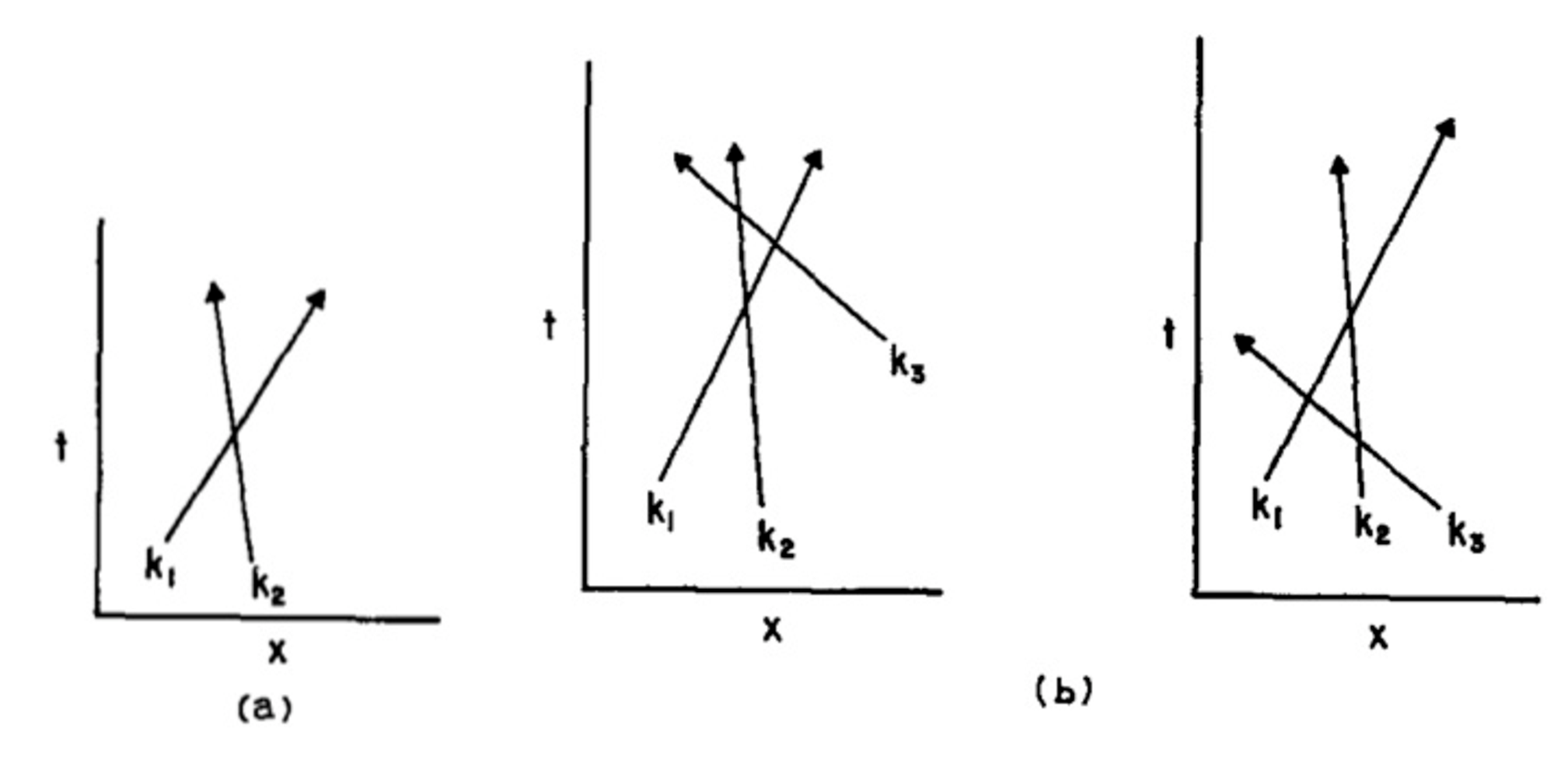}
\end{center}
\caption{Space-time plots for (a) two- and (b) three-particle problems.  
The point is that the two factorisations in (b) are equivalent. From McGuire.\cite{McGuire1}}
\label{Mcguirefig}
\end{figure}

There is a deep reason for why these models are integrable. 
At the heart is the Yang-Baxter relation, which has appeared in many guises.\footnote{See, e.g., 
the various books, review articles and lecture notes in 
Refs. \refcite{Baxter,Gaudin_book,TW,Sutherland,Perk,Korepin,Schlottmann,Sutherland_book,Mussardo,McCoy}, 
which is by no means a complete list. Indeed, the brief overview given in the present article is necessarily incomplete. 
In some sense this article is a sequal to Ref. \refcite{Batch}.}
Our interest here is in the context of quantum many-body systems for which the 
key ingredient is the scattering matrix ($S$-matrix).
In particular, our interest is in models for which the $S$-matrix of an $N$-particle system  
factorizes into a product of $N(N-1)/2$ two-body $S$-matrices. 
For models confined to one space dimension this factorisation is represented as 
a space-time scattering diagram in \fref{Mcguirefig}.

The condition of integrability is equivalent to a condition of no diffraction.\cite{McGuire1,Sutherland_book} 
Indeed, the Bethe Ansatz can only be applied when there is non-diffracting scattering.
The notions of non-diffracting scattering and quantum integrability are essentially equivalent in the 
present context.
This is one of the best available definitions of quantum integrability.\cite{Caux}

In the next section the above examples will be discussed briefly.

\section{Factorized scattering and key integrable models} 

In a completely integrable system the three-body $S$-matrices corresponding to the two diagrams in \fref{Mcguirefig}(b)  
are equal and have the factorisation equation 
\begin{equation}
S(1,2,3) = S(2,3) S (1,3) S(1,2) = S(1,2) S(1,3) S(2,3) 
\end{equation}
where $S(i,j)$ is the two-body $S$-matrix acting on states $i$ and $j$.
In the Yang-Baxter language, this form of $S(i,j)$ is identical to Yang's operator $X_{ij} = P_{ij} Y^{ij}_{ij}$, or
equivalently, to Baxter's $R$-matrices.\footnote{Yang's masterstroke was to translate McGuire's geometric 
construction into operator form.\cite{Yang1,Yang2}} 
 
\subsection{bosons}

The hamiltonian of $N$ interacting spinless bosons on a line of length $L$ ($\hbar =2m=1$) 
with point interactions is 
\begin{equation}
 {\cal H}=-\sum_{i = 1}^{N}\frac{\partial ^2}{\partial x_i^2}+2\,c \sum_{1\leq i<j\leq N}\delta (x_i-x_j)
\label{Hamb}
\end{equation}
where $x_i$ are the boson co-ordinates and $c$ is the interaction strength. This is the model solved in Lieb and Liniger\cite{Lieb} by means 
of the Bethe Ansatz wavefunction\footnote{The amplitudes $A(P)$ involve a sum over permutations 
$P=(P_1,\ldots,P_N)$ of $(1,\ldots,N)$.}
\begin{equation}
\psi(x_1,\ldots, x_N)=\sum_{P}A(P) \exp(\mathrm{i}\sum_{j=1}^Nk_{P_j}x_j)
\end{equation} 
which gives the energy eigenvalues 
\begin{equation}
{\cal E}=\sum_{j=1}^N k_{j}^2  
\label{En}
\end{equation}
in terms of the roots $k_j$ of Bethe equations of the form
\begin{equation}
\exp(\mathrm{i}k_jL)=- \prod^N_{\ell = 1} 
\frac{k_j-k_\ell+\mathrm{i}\, c}{k_j-k_\ell-\mathrm{i}\, c} \qquad \mbox{for} \quad j = 1,\ldots, N  \,.
\end{equation}
This is arguably the simplest set of known Bethe equations, for which all of the roots $k_j$ are real 
in the repulsive regime $c>0$.

For this model the two-body $S$-matrix element for $k_1 < k_2$ is $S(k_2,k_1) = S(k_2-k_1) = S(p)$, where 
\begin{equation}
S(p) = \frac{p - {\rm i} c}{p + {\rm i} c} 
\end{equation}
with $p$ the rapidity and $S(p) S(-p) =1$.

\begin{figure}[h]
\begin{center}
\includegraphics [width=0.70\linewidth]{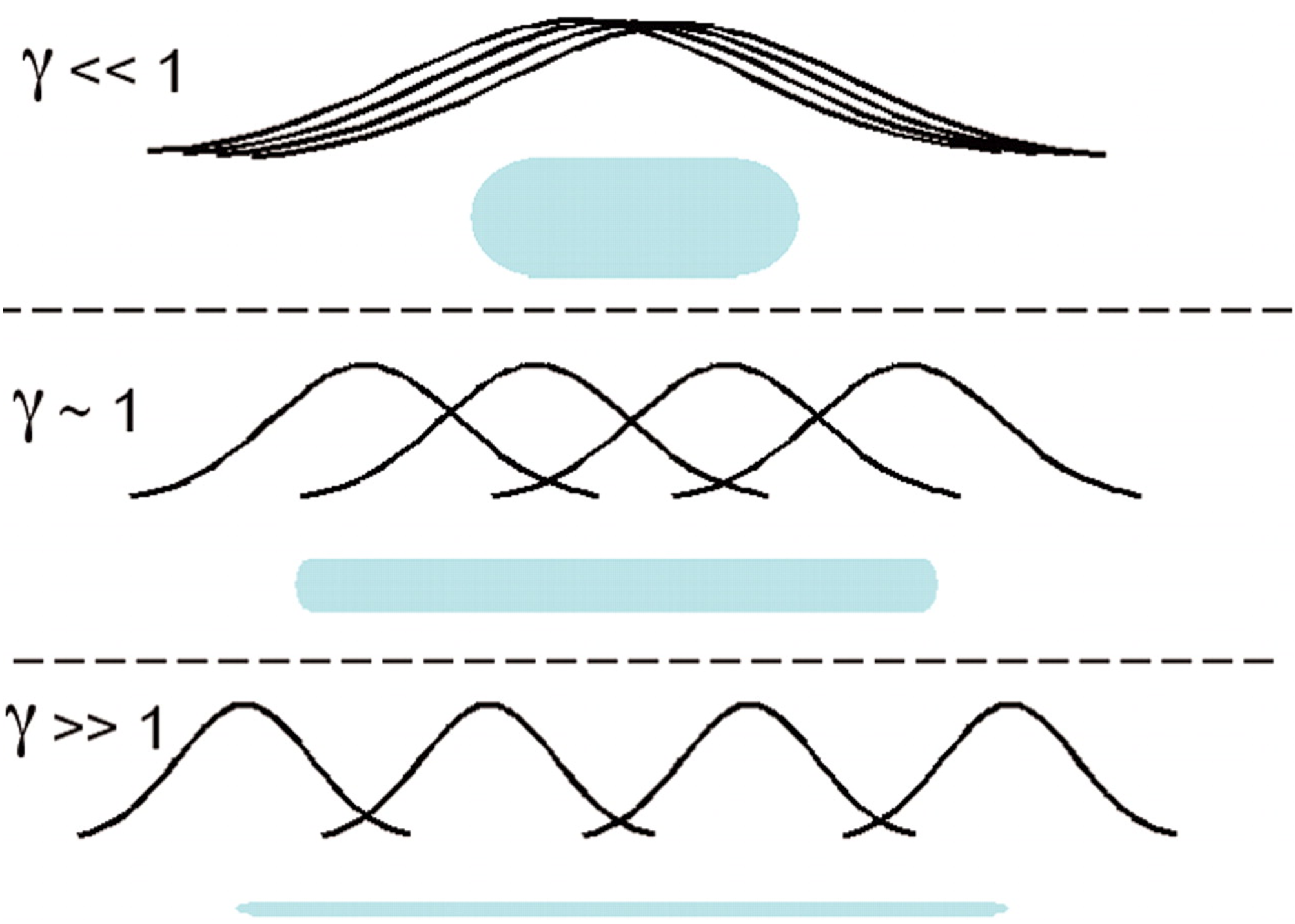}  
\end{center}
\caption{Cartoon showing the `fermionisation' of bosons as the interaction strength $\gamma$ is increased. 
For $\gamma \ll 1$ the behaviour is like a condensate, whereas for $\gamma \gg 1$ the behaviour is like the hard-core 
bosons of the Tonks-Girardeau gas. From Kinoshita, Wenger and Weiss.\cite{KWW}
}
\label{cartoonfig}
\end{figure}

In the analysis of this model it is convenient to define the dimensionless interaction parameter 
$\gamma = c/n$ in terms of the number density $n=N/L$.  
A cartoon of the atom distributions, representing the `fermionisation' of the one-dimensional interacting Bose gas 
with increasing $\gamma$ is shown in \fref{cartoonfig}.

The attractive regime $c<0$ has also been of interest. Inspired by Monte Carlo results which  
predicted the existence of a super Tonks-Girardeau gas-like state in the attractive interaction 
regime of quasi-one-dimensional Bose gases,\cite{Astra} it was shown that 
a super Tonks-Giradeau gas-like state corresponds to a highly-excited Bethe state in the integrable Bose gas 
with attractive interactions, for which the bosons acquire hard-core behaviour.\cite{sTG}
The large kinetic energy inherited from the Tonks-Girardeau gas -- as the interaction is switched from strongly repulsive to strongly attractive -- 
in a Fermi-pressure-like manner, prevents the gas from collapsing.

\subsection{fermions}

The hamiltonian of the one-dimensional fermion problem is similar to \eref{Hamb}, with
\begin{equation}
 {\cal H}=-\sum \frac{\partial ^2}{\partial x_i^2} - \sum \frac{\partial ^2}{\partial y_i^2}+2\,c \sum \delta (x_i-y_j)
\label{Hamf}
\end{equation}
where $x_i$ and $y_i$ are the co-ordinates of the spin-up and spin down fermions.
In this case there are a total of $N$ interacting two-component fermions on a line of length $L$, 
with $M$ the number of spin-down fermions. 
The energy expression is also the same as \eref{En}. 
However, because spin is involved, the Bethe equations are now of the more complicated -- nested -- form\cite{Gaudin,Gaudin_thesis,Yang1} 
\begin{eqnarray}
\exp(\mathrm{i}k_jL)&=&\prod^M_{\ell = 1} 
\frac{k_j-\Lambda_\ell+ \frac12\, \mathrm{i}\,c}{k_j-\Lambda_\ell-\frac12\, \mathrm{i}\,c}\\
\prod^N_{\ell = 1}\frac{\Lambda_{\alpha}-k_{\ell}+\frac12 \,\mathrm{i}\,c}
{\Lambda_{\alpha}-k_{\ell}-\frac12\, \mathrm{i}\,c}
 &=& - {\prod^M_{ \beta = 1} }
\frac{\Lambda_{\alpha}-\Lambda_{\beta} +\mathrm{i}\, c}{\Lambda_{\alpha}-\Lambda_{\beta} -\mathrm{i}\, c} 
\end{eqnarray}
for $j = 1,\ldots, N$ and $\alpha = 1,\ldots, M$.
We may define the polarization by $P = (N-2M)/N$. 
The special case $M=N/2$ for which $P=0$ is known as the balanced case.

The matrix form of the associated wavefunction and $S$-matrix are   
not discussed here, rather the reader is referred 
to the literature\cite{Gaudin_book,Sutherland_book,GBL}. 

\begin{figure}[h]
\begin{center}
\includegraphics[width=0.2\linewidth,angle=-90]{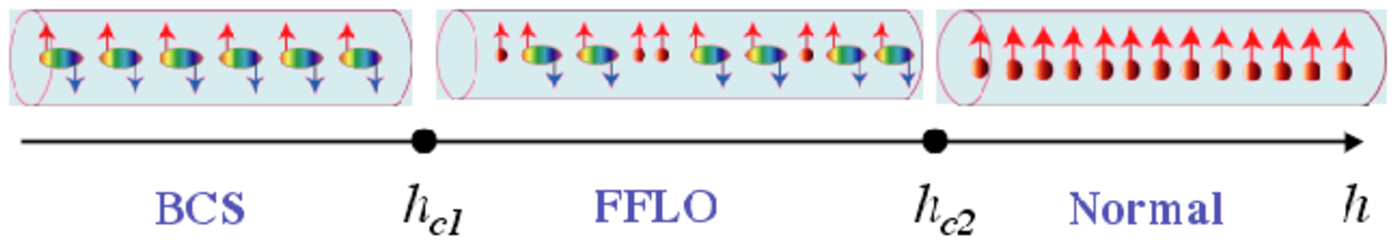}  
\end{center}
\caption{This figure is essentially the zero temperature phase diagram of the Gaudin-Yang model 
as a function of magnetic field $h$ for given chemical potential. The three phases are the fully paired (BCS) phase, which is a quasi-condensate 
with zero polarization ($P = 0$), the fully polarized (Normal) phase with $P = 1$, and the partially polarized (FFLO) phase where $0 < P < 1$. 
The FFLO phase can be viewed as a mixture of pairs and leftover (unpaired) fermions. 
For given chemical potential, the FFLO phase is separated from the BCS phase and the normal phase by the quantum critical points 
$h_{c1}$ and $h_{c2}$. From Zhao and Liu.\cite{ZL}
}
\label{schematic}
\end{figure}

\begin{figure}[h]
\begin{center}
\includegraphics [width=0.55\linewidth,angle=-90]{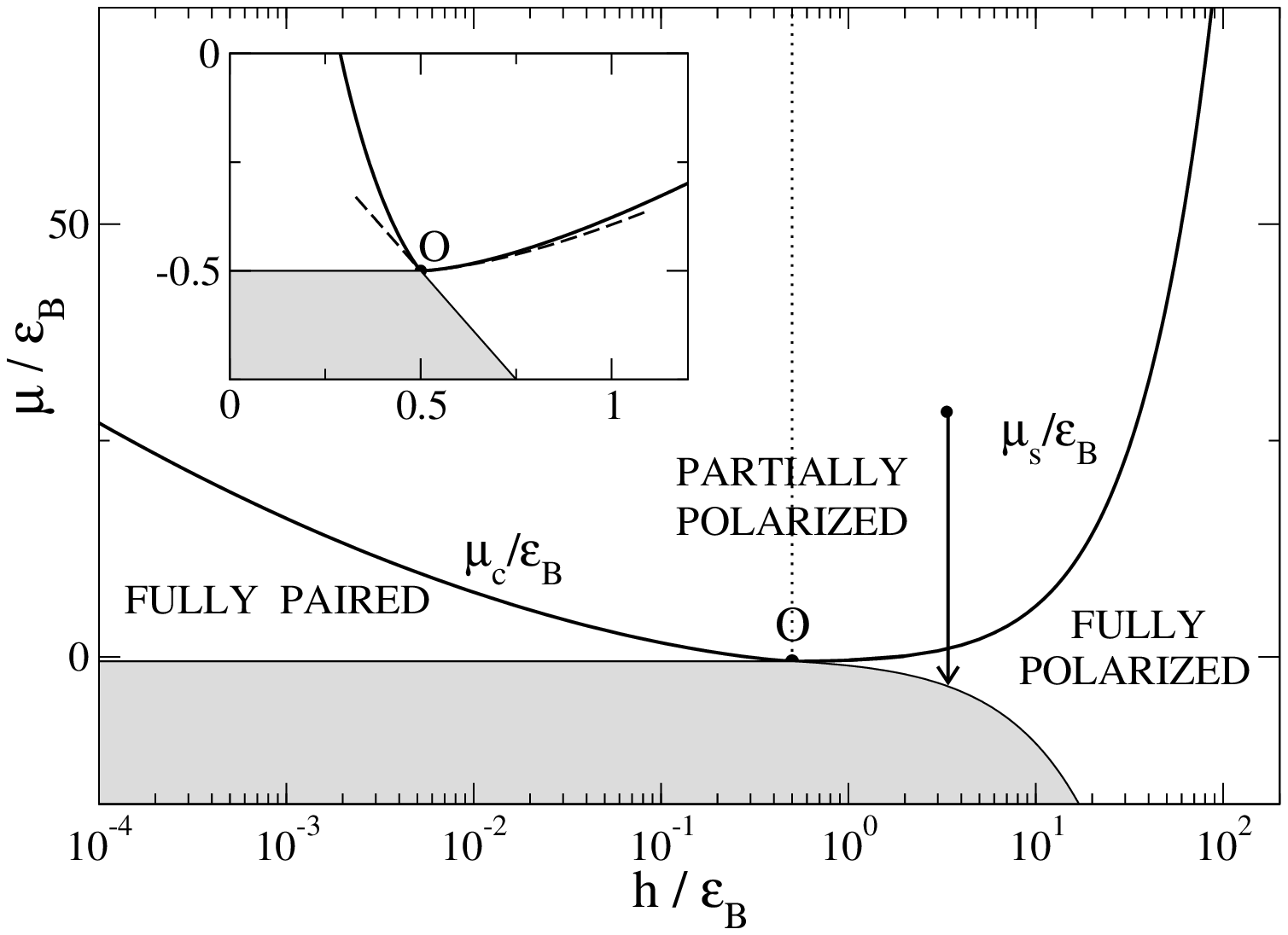}  
\end{center}
\caption{Phase diagram of the Gaudin-Yang model as a function of chemical potential and magnetic field obtained 
from using numerical solution of the Bethe equations. From Orso.\cite{Orso}
}
\label{orsofig}
\end{figure}

In the attractive regime the Bethe roots tend to form pairs which can be broken by applying a magnetic field to the hamiltonian \eref{Hamf}. 
The quantum critical points distinguishing the different quantum phases (see \fref{schematic}) can be calculated analytically 
and the full phase diagram mapped out (see \fref{orsofig}).\cite{GBL}

Much has been written about the one-dimensional Hubbard model,\cite{Wu} which 
is also of fundamental importance.\cite{H-book}
The model describes interacting electrons in narrow energy bands and 
in the continuum limit is equivalent to the interacting two-component fermion model. 

\subsection{E$_8$ and the Ising model}

Like the result of the alluring call of the Sirens in Greek mythology, attempts to solve the two-dimensional 
classical Ising model in a magnetic field -- and its one-dimensional quantum counterpart -- 
have foundered on the rocks. Fortunately there is a modern day Orpheus. 
Zamolodchikov\cite{Zam} discovered a remarkable integrable quantum field theory 
containing eight massive particles with a reflectionless factorized $S$-matrix.
This is the $c = \frac12$ conformal field theory (corresponding to the critical Ising model) 
perturbed with the spin operator $\phi_{1,2} = \phi_{2,2}$ of dimension ($\frac{1}{16}, \frac{1}{16}$).
Up to normalisation, the masses $m_i$ of these particles coincide with the 
components $S_i$ of the Perron-Frobenius vector of the 
Cartan matrix of the Lie algebra $E_8$:  $m_i/m_j = S_i/S_j$.

With normalisation $m_1=1$, the masses\footnote{Note the appearance of the golden ratio for $m_2/m_1$.} 
are\cite{Zam} 
\begin{equation*}
\begin{array}{ll}
m_2 = 2 \cos \frac{\pi}{5}  & = 1.618~033\ldots \\
m_3 = 2 \cos \frac{\pi}{30} & = 1.989~043\ldots \\
m_4 = 4 \cos \frac{\pi}{5} \cos \frac{7\pi}{30} & = 2.404~867\ldots \\
m_5 = 4 \cos \frac{\pi}{5} \cos \frac{2\pi}{15} & = 2.956~295\ldots \\
m_6 = 4 \cos \frac{\pi}{5} \cos \frac{\pi}{30}  & = 3.218~340\ldots \\
m_7 = 8 \cos^2 \frac{\pi}{5} \cos \frac{7\pi}{30} & = 3.891~156\ldots \\
m_8 = 8 \cos^2 \frac{\pi}{5} \cos \frac{2\pi}{15} & = 4.783~386\ldots
\end{array}
\end{equation*}

The $S$-matrix of this model is particularly impressive. 
The $S$-matrix element describing the scattering of the lightest particles is given by \cite{Zam}
\begin{equation}
S_{1,1}(\beta) = \frac{\tanh \left(\frac{\beta}{2}+{\rm i} \frac{\pi}{6}\right) 
\tanh \left(\frac{\beta}{2}+{\rm i} \frac{\pi}{5}\right)
\tanh \left(\frac{\beta}{2}+{\rm i} \frac{\pi}{30}\right)}
{\tanh \left(\frac{\beta}{2}-{\rm i} \frac{\pi}{6}\right) \tanh \left(\frac{\beta}{2}-{\rm i} \frac{\pi}{5}\right) 
\tanh \left(\frac{\beta}{2}-{\rm i} \frac{\pi}{30}\right)}
\end{equation}
where $\beta$ is the rapidity. The other elements are 
uniquely determined by the bootstrap program.

The $E_8$ theory is conjectured to describe the scaling limit of the two-dimensional classical Ising model 
in a magnetic field.\footnote{For recent work with regard to scaling and universality, 
see Ref. \refcite{Dudalev} and references therein.}
The first several masses were soon confirmed numerically for the one-dimensional quantum counterpart, 
the quantum Ising chain with transverse and longitudinal fields.\cite{Henkel}
A realisation exists in terms of the dilute $A_3$ lattice model \cite{Nienhuis} -- 
an exactly solved lattice model in the same universality class as the 
two-dimensional Ising model in a magnetic field -- 
from which the $E_8$ mass spectrum has been derived.\cite{Bazhanov,Seaton}

As we shall see further below, the fact that the emergence of such an exotic symmetry 
as $E_8$ can be observed in the lab is quite remarkable.

\section{Experiments}

In this section a brief sketch is given of experiments which have made contact with the models 
discussed above.\footnote{The particular experiments chosen are selective and by no means exhaustive.}
These experiments have been a result of the ongoing `virtuoso triumphs' in experimental techniques -- 
in cold atom optics and in the more traditional setting of condensed matter physics.
As a result it is now possible to probe and understand the physics of key quantum many-body systems which 
should ultimately be of benefit to quantum technology. 

\subsection{bosons}

Experiments on the trapping and cooling of bosonic atoms in tight one-dimensional waveguides and related 
theoretical progress have been recently reviewed.\cite{Olshani,Caz}
Most importantly, it is possible to confine atoms to effectively one-dimensional tubes and to vary the interaction strength 
between atoms, both in the repulsive and attractive regimes.

\begin{figure}[h]
\begin{center}
\includegraphics [width=0.80\linewidth]{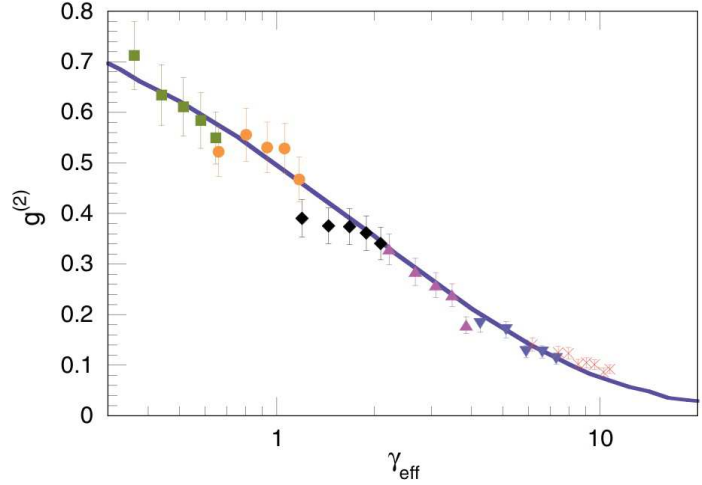}  
\end{center}
\caption{Local pair correlation function $g^{(2)}$ obtained from photoassociation rates vs effective interaction strength. 
The solid line is obtained from the Lieb-Liniger model. From Kinoshita, Wenger and Weiss.\cite{KWWg2}
}
\label{g2fig}
\end{figure}

One of the early experiments which made contact with the one-dimensional Lieb-Liniger model of interacting bosons 
measured local pair correlations in bosonic Rb atoms by photoassociation. 
The local pair correlation function $g^{(2)}$ is proportional to the probability of observing two 
particles in the same location. The experimental measurement of $g^{(2)}$ by 
Kinoshita, Wenger and Weiss\cite{KWWg2} is shown in \fref{g2fig}. 
As expected, the curve drops off towards zero as the interaction strength increases, just like in a 
non-interacting Fermi gas (recall \fref{cartoonfig}).

\begin{figure}[h]
\begin{center}
\includegraphics [width=0.98\linewidth]{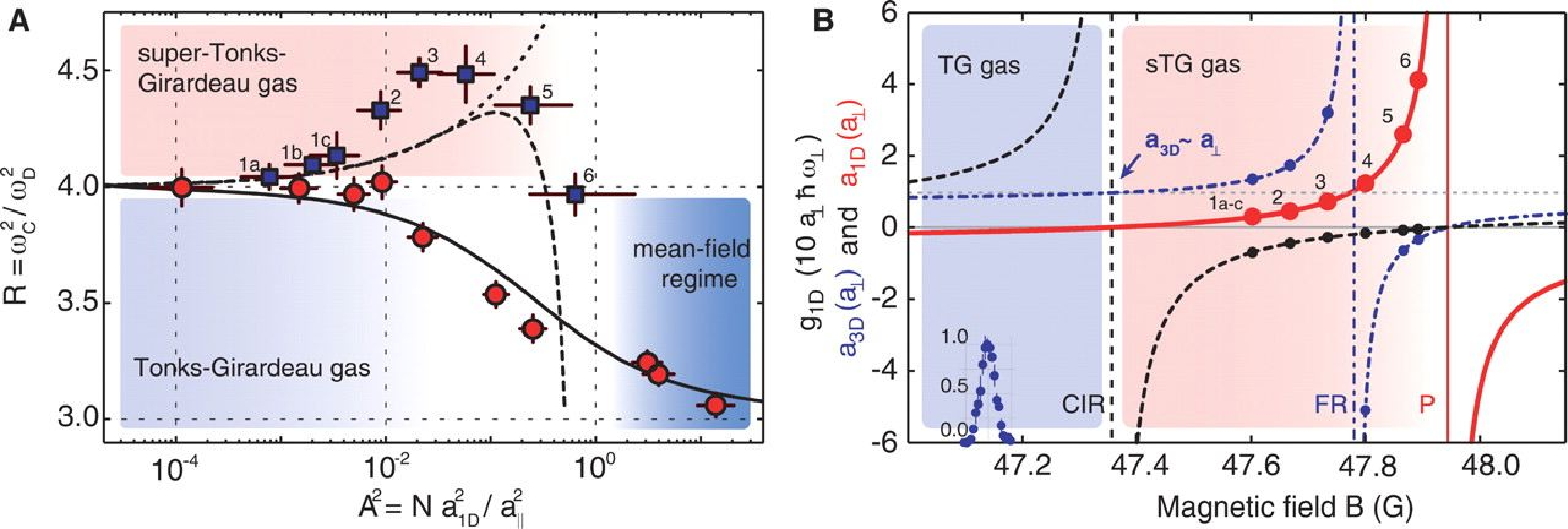}  
\end{center}
\caption{Plots of experimental data providing evidence for the super Tonks-Girardeau gas. From Haller {\em et al.}\cite{Haller}. 
}
\label{sTGfig}
\end{figure}

Experiments have also been performed on one-dimensional bosons in the attractive regime.
In particular, using a tunable quantum gas of bosonic cesium atoms, Haller {\em et al.}\cite{Haller} realized and controlled in 
one-dimensional geometry a highly excited quantum phase -- the super Tonks-Girardeau gas -- that is stabilized in the 
presence of attractive interactions by maintaining and strengthening quantum correlations across a confinement-induced resonance (see \fref{sTGfig}).
They diagnosed the crossover from repulsive to attractive interactions in terms of the stiffness and energy of the system. 
This opened up the experimental study of metastable, excited, many-body phases with strong correlations.

\subsection{fermions}

\begin{figure}[h]
\begin{center}
\includegraphics [width=0.8\linewidth]{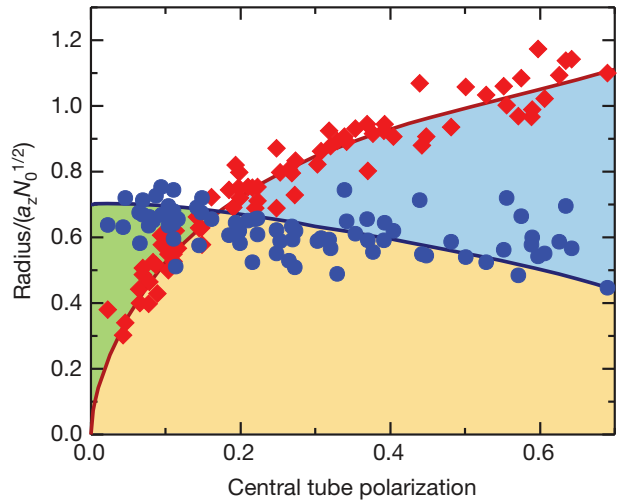}  
\end{center}
\caption{Experimental phase diagram of one-dimensional two-component fermions as a function of polarization. 
The red diamonds and blue circles denote the scaled radii of the axial density difference and the minority state axial density, 
respectively. The solid lines follow from the Gaudin-Yang model. From Liao {\em et al.}\cite{Rice}
}
\label{fermionfig}
\end{figure}

Theoretical progress and experiments on fermionic atoms confined to one-dimension have been recently reviewed.\cite{GBL} 
Of particular relevance here is the experiment performed at Rice University using fermonic $^6$Li atoms.\cite{Rice} 
The system has attractive interactions with a spin population imbalance caused by a difference in the number of spin-up and spin-down atoms. 
Experimentally, the gas is dilute and strongly interacting.
The key features of the phase diagram (recall \fref{orsofig}) have been experimentally confirmed using 
finite temperature density profiles (see \fref{fermionfig}).\cite{Rice}
The system has a partially polarized core surrounded by either fully paired or fully polarized wings at low temperatures, 
in agreement with theoretical predictions.\cite{GBL}
More generally, this work experimentally verifies the coexistence of pairing and polarization at quantum criticality.  

In further developments, experiments have also been performed with just two distinguishable $^6$Li atoms.\cite{Zurn1}
This provides an experimental study of one-dimensional fermionisation as a function of the interaction strength. 
For a magnetic field below the confined induced resonance two interacting fermions form a Tonks-Girardeau state 
whereas a super Tonks-Girardeau gas is created when the magnetic field is above the resonance value. 
Quasi one-dimensional systems consisting of up to six ultracold fermionic atoms in two different spin states with 
attractive interactions have also been studied experimentally,\cite{Zurn2} 
including the crossover from few to many-body physics.\cite{Wenz}

\subsection{E$_8$ and the quasi-1D Ising ferromagnet CoNb$_2$O$_6$}

\begin{figure}[ht]
\begin{center}
\includegraphics [width=0.85\linewidth]{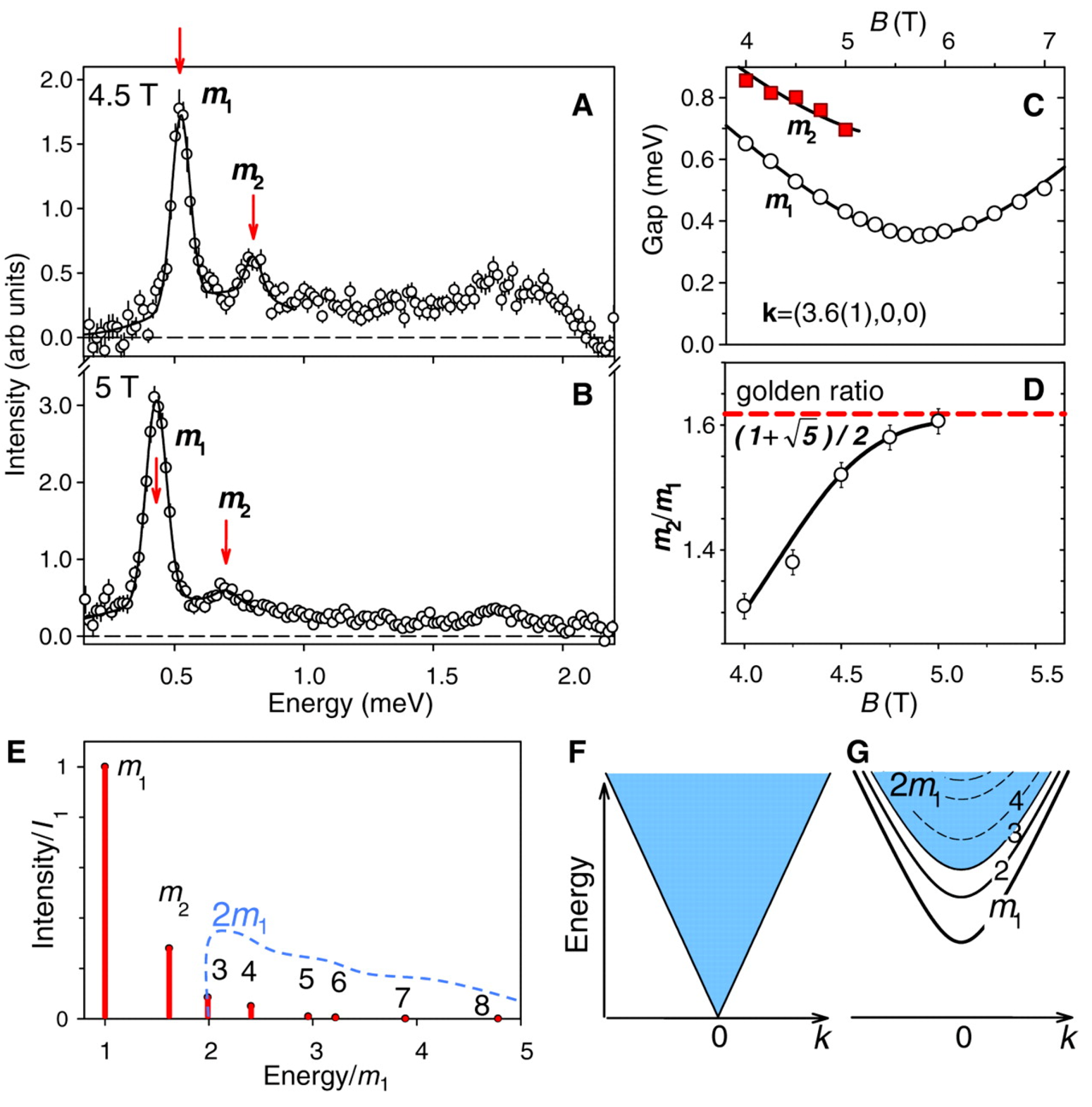}  
\end{center}
\caption{Various plots showing experimental evidence for the first few masses of 
the $E_8$ mass spectrum in the transverse Ising chain. 
From Coldea {\em et al.}\cite{Coldea}. 
}
\label{massfig}
\end{figure}

In an experiment in the traditional setting of condensed matter physics, Coldea {\it et al.}\cite{Coldea} realised a 
quasi-one-dimensional Ising ferromagnet in CoNb$_2$O$_6$ (cobalt niobate) tuned through its quantum critical point 
using strong transverse magnetic fields. 
The underlying Ising hamiltonian 
\begin{equation}
H = - J \sum_i s^z_i s^z_{i+1} - h \, s_i^x - h_z \, s_i^z
\end{equation}
has a quantum critical point at $h=h_c = J/2$ for $h_z=0$. 
In the scaling limit sufficiently close to the quantum critical point, i.e., $h_z \ll J, h=h_c$, 
the spectrum is predicted to be described by Zamolodchikov's $E_8$ mass spectrum.
Coldea {\em et al.}\cite{Coldea} were able to observe the spectrum by neutron scattering. 
In particular, they were able to observe evidence for the first few $E_8$ masses, see \fref{massfig}.

In fact the integrable theory provides many more exact predictions than
experiments have been able to test so far,\cite{aldo} involving, for example, correlation functions.\cite{dm}
Since the work of Coldea {\em et al.}\cite{Coldea} it is reasonable to expect further progress on the experimental side.

\section{Concluding remarks}

We have seen through the few examples given here that the twin concepts of non-diffractive  
and factorized scattering -- embodied in the Yang-Baxter equation -- have captured the fundamental physics 
of some key interacting quantum many-body systems. 
It could hardly have been imagined in the 1960's that such mathematical models would some day 
make contact with experiment.
The philosophy and pioneering spirit of the 1960's was 
captured at that time in the compilation of introductory material and original articles\footnote{
Including material and a reprinted paper by F.~J. Dyson on the dynamics of disordered chains. 
On a personal note, we are all familiar with Dyson's early mastery of the $S$-matrix in quantum electrodynamics  
and his brilliant synthesis of the different formulations of quantum electrodynamics 
due to Tomonaga, Schwinger and Feynman. My favourite Dyson moment is his legendary common room encounter
with Hugh Montgomery in the early 1970's which set alight the discovery of the 
remarkable and deep connection between the distribution of 
zeros of the Riemann zeta function and random Hermitian matrices. Dyson's perspective, of course, was from his earlier 
seminal work on the statistical features of the level spacings of quantum systems.
} 
in the book  
{\em Mathematical Physics in One Dimension} by Lieb and Mattis.\cite{LM}
Over the following decades further theoretical progress and the striking developments in experimental technology have 
revealed that physics in one dimension is indeed a particularly rich and worthwhile pursuit, \cite{Giamarchi} 
and {\em does} provide a path to understanding nature.

There has long been a school of thought, with which Professor Dyson concurs, 
that mathematical models should only be tackled in earnest if there is a prospect that  
some day they may be relevant to experiments.  
Yet we have seen from the examples of the one-dimensional Bose and Fermi gases that it 
may take up to, and even more than, 40 years before mathematical models of this kind move out of the paper and into the lab. 
Fortunately this has been during the lifetime of those involved in the pioneering developments of the 1960's and later.

\section*{Acknowledgments}

The author gratefully acknowledges support from the 1000 Talents Program of China and from Chongqing University. 
This work has also been supported by the Australian Research Council. 
The author also thanks John Cardy, All Souls College and the Rudolf Peierls Centre for Theoretical Physics 
for kind hospitality and support during his visit to Oxford. It is a pleasure to thank Rodney Baxter, Jean-S\'ebastien Caux, 
Aldo Delfino, Fabian Essler, Angela Foerster, Xiwen Guan, Austen Lamacraft,  
Giuseppe Mussardo and Huan Zhou for useful discussions during the course of writing this article.

\end{document}